\documentstyle[fleqn,espcrc2,psfig]{article}

\title{The Complex Structure of 2D Surfaces} 
\author{H.Kawai, N.Tsuda and T.Yukawa \\
National Laboratory for High Energy Physics (KEK), Tsukuba 305, Japan
}

\begin{document}
\begin{abstract}
The complex structure of a surface generated by the two-dimensional dynamical 
triangulation(DT) is determined by measuring the resistivity of the surface.
It is found that surfaces coupled to matter fields have well-defined complex 
structures for cases when the matter central charges are less than or equal 
to one, while they become unstable beyond $c = 1$.
A natural conjecture that fine planar random network of resistors behave as a 
continuous sheet of constant resistivity is justified numerically for $c<1$.
\end{abstract}
\maketitle

\section{Introduction}
The discretized approach of 2D quantum gravity, implemented by the matrix 
model, exhibits, in a continuum limit the same behavior as the continuous 
approach, given by the Liouville field theory.
Thus it is certain that the two theories describe the same physics.
However, we feel that the mutual relationship is not yet fully understood.
While the relation between DT and matrix model is transparent, its relation 
to the Liouville field theory is not so evident.
Intuitively, we expect that the simplicial manifolds generated by DT will 
tend to continuous manifolds in the large number limit of simplexes.
However, such surfaces are known to be fractal\cite{fractal1}, hence the way 
the lattice regularization simulates the continuous theory is not at all clear. 
It is therefore indispensable to make sure that an universal continuum limit 
exists.
In order to check this numerically, we propose to measure how the network 
conducts currents, and compare it with the corresponding continuous medium.
An invariance under local scale transformation plays an essential role in the 
algorithm, and because of this property we can extract information about the 
complex structure which is independent of the conformal mode parametrised by 
moduli.

Despite many theoretical developments the precise nature of the transition 
at $c = 1$ is still not fully understood.
Serious numerical efforts have failed so far to convincingly establish the 
transition, at least by measuring the string susceptibility.
It should be much easier to observe this transition in 
local scale invariant quantities, rather than in the original metric which
is comparatively complicated reflecting the fractal nature of the surface.
Here, we show such a method by which we can define the complex 
structure of a DT surface and separate the conformal mode.

\section{Determination of the complex structure}
The basic idea of our method originates from the analog model 
\cite{fishnet_analog} of the dual amplitude in analogy with the electrical 
circuit\cite{circuit}.
Let us first consider a two dimensional conducting medium with conductivity 
tensor density $\sigma^{\mu \nu}$.
Ohm's law reads
\begin{equation}
j^{\mu}= \sigma^{\mu \nu} E_{\nu},
\label{eq:Ohm_law}
\end{equation}
where $E_{\nu}(=-\partial_{\nu}V)$ is an electric field and $j^{\mu}$ is a 
current density.
The Joule heat $Q$ can be estimated by a potential distribution $V$ on the 
surface; $Q=\int d^2x \sigma^{\mu \nu} \partial_{\mu}V \partial_{\nu}V$.
Requiring $\delta Q=0$, it leads to an equation for $V$: 
$\partial_{\mu}\sigma^{\mu \nu} \partial_{\nu}V = 0$, which is just the 
equation of continuity.
For a homogeneous isotropic medium the conductivity tensor density is given by
$\sigma^{\mu \nu}= \frac{1}{r} \sqrt{g} g^{\mu \nu}$ with $r$ being the 
resistivity constant.  
Therefore, Ohm's law(\ref{eq:Ohm_law}) is invariant under the local scale 
transformation $g_{\mu \nu} \mapsto  g_{\mu \nu} e^{-\sigma}$.
In the following we first give an algorithm to measure the resistivity from 
voltage distribution.
In the case of spherical topology we can regard the surface essentially as an 
infinite flat sheet. 
Then the potential at a point specified by the complex coordinate, 
$z$, with a source of current $I$ placed at $z_{k}$, and a sink 
of the current at $z_{l}$ is given by 
\begin{equation}
V(z)=-{Ir \over 2\pi} \ln  | \frac{z-z_{k}}{z-z_{l}} | + {\rm Const}.,
\label{eq:V(z)}
\end{equation}
In order to avoid the ambiguity arising from Const. in eq.(\ref{eq:V(z)}) 
we measure the potential drop between coordinates $z_{i}$ and $z_{j}$. 
We define the potential drop between points $i$ and $j$ in the presence of a 
source and a sink of current at points $k$ and $l$ by 
\begin{eqnarray}
V_{ij}^{(k)(l)} 
&\equiv& V(z_{i})-V(z_{j}) \nonumber \\
&=& -{Ir \over 2\pi} \ln |[z_i,z_j;z_k,z_l]|,
\label{eq:potential}
\end{eqnarray}
where $[z_i,z_j;z_k,z_l]={z_i - z_k \over z_i - z_l}
{z_j - z_l \over z_j - z_k}$ is known as the anharmonic ratio.
It has an important property: {\it i.e.} $SL(2,C)$ invariance.
Then it allows us to fix three coordinates among the $\{z_i,z_j,z_k,z_l\}$ to 
any desired values without changing the potential drop eq.(\ref{eq:potential}).
For example, we can fix three coordinates as $z_{j}=1$, $z_{k}=0$, and 
$z_{l}=\infty$. 
Then the potential drop eq.(\ref{eq:potential}) for $I=1$ is written as
\begin{equation}
V_{ij}^{(k)(l)}=-{r \over 2\pi} \ln | z_{i} |,
\label{eq:potd1}
\end{equation}
\begin{equation}
V_{ik}^{(j)(l)}=-{r \over 2\pi} \ln | 1-z_{i} | .
\label{eq:potd2}
\end{equation}
All the other possible permutations of $\{z\}$ only give 
the potential drops obtained by linear combinations of eq.(\ref{eq:potd1}) 
and eq.(\ref{eq:potd2}). 
Since unknown variables are $\{z_{i},r \}$, we need the potential drop at an 
additional point 
$z_{m}$.
Then we get three more relations in addition to eq.(\ref{eq:potd1}) and 
eq.(\ref{eq:potd2}) which altogether are sufficient to specify five unknowns: 
$\{z_{i},z_{m},r \}$. 
The three extra equations are given by
\begin{equation}
V_{mj}^{(k)(l)}=-{r \over 2\pi} \ln | z_{m} |,
\end{equation}
\begin{equation}
V_{mk}^{(j)(l)}=-{r \over 2\pi} \ln | 1-z_{m} |,
\end{equation}
\begin{equation}
V_{ik}^{(m)(l)} + V_{mj}^{(k)(l)}=-{r \over 2\pi} \ln | z_{i} - z_{m} |.   
\end{equation}
Now we apply this to a random surface generated by DT.
Since the local scale invariance is not fully realized on this surface, 
the resistivity may have certain fluctuations.
In fact, the value varies from one sample to another, and depending on 
the position of electrodes for the measurements.
Such ambiguities should, however, vanish in the continuum limit, if it exists 
at all. 
The dual graph of a surface consisting of $N$ triangles is regarded as a 
trivalent network, where we fix the resistance of the link connecting two 
neighbouring vertices to be $1$.
What we want to do is to examine whether such a network behaves like a 
continuous medium in the limit of $N \rightarrow \infty$.
For the measurement we pick five vertices in the dual graph and perform 
the method as explained above.
The practical method we employ for the determination of potential drops
is as follows;
we pick two vertices at $v_{in}$ and $v_{out}$ for the source and the sink 
of current with unit intensity.
By writing the potential of the vertex $v$ as $V(v)$, current conservation 
reads $V(v)=\frac{1}{3} \{ \sum_{i=1,2,3} V(v_{i}) + \delta_{v,v_{in}} 
- \delta_{v,v_{out}} \},$ where $v_{1},v_{2}$ and $v_{3}$ are the three 
neighboring vertices of $v$.
We solved this set of equations by the {\it successive overrelaxation} method.

\section {Analog model}
Our method can be reconsidered from the string-theoretical point of view.
Whether string theory can be described by local fields or not is one of 
the fundamental questions.
The authors of \cite{fishnet_analog} showed that a sum of planar fishnet 
Feynman diagrams with $N$ external particles is indeed approximated by a 
$N$ particle Veneziano amplitude.
The basic assumption of this model is that the network corresponding to 
a fine planar fishnet Feynman diagram can be regarded as a homogeneous 
isotropic conducting sheet with a constant resistivity.
Once this is accepted it is straightforward to derive dual amplitudes
by evaluating the heat generated by the sheet.
The purpose of our article is to justify this basic assumption 
numerically.
In order to clarify the correspondence to the network of resistors, 
we write the amplitude of fishnet Feynman diagrams as 
\begin{equation}
\mathop{\sum}_{\stackrel{\mbox{\scriptsize planar}}
{\mbox{\scriptsize diagrams}}} \! \int \!\! 
\mathop{\prod}_{l:\mbox{\scriptsize propagator}} 
\!\! d\alpha_{l} \rho(\alpha_{l}) \int \! \prod_{i:\mbox{\scriptsize vertex}} 
d^{D} x_{i} e^{-Q}
\end{equation}
where $Q = \sum_{l} \frac{(x_{i}-x_{j})^{2}}{2 \alpha_{l}}$, 
$l$ and ($i,j$) indicate a link and its two ends(vertices) of the 
network respectively, and $\rho(\alpha_{l}) = \frac{1}{2(2\pi)^{D/2}} 
\alpha_{l}^{-\frac{D}{2}} e^{- \frac{1}{2} \alpha_{l} m^{2}}.$
Here we have introduced a Feynman parameter $\alpha_{l}$ 
\footnote{
There are analogies between Feynman diagram and network of resistors:
($\alpha_{l} \leftrightarrow$ resistance), 
($x_{i} \leftrightarrow$ voltage at i), 
($p_{l} \leftrightarrow$ current through l) and 
($Q \leftrightarrow$ heat generation).
}
for each propagator $l$ in the 
momentum integration, 
\begin{eqnarray*}
\lefteqn{\int \frac{d^{D}p}{(2\pi)^{D}} \frac{1}{p^{2}+m^{2}}e^{ip(x-y)}}\\
&& = \int_{0}^{\infty} d\alpha_{l} \frac{1}{2(2\pi)^{D}} 
\alpha_{l}^{-\frac{D}{2}} e^{-\frac{1}{2} \alpha_{l} m^{2}}
e^{-\frac{(x-y)^{2}}{2\alpha_{l}}}.
\end{eqnarray*}

The basic conjecture is that the fishnet diagrams behaves as a continuum 
homogeneous conducting sheet after taking the average 
$ 
\mathop{\sum}_{\stackrel{\mbox{\scriptsize planar}}
{\mbox{\scriptsize diagrams}}} \! \int \!
\mathop{\prod}_{l:\mbox{\scriptsize propagator}} 
\! d\alpha_{l} \rho(\alpha_{l}). 
$

\section {Numerical results and discussions}
We fix the topology to be two sphere($S^2$) and the number of triangles 
to be $4,000$, $8,000$ and $16,000$ (prohibiting tadpole and self-energy 
diagrams).
For matter fields coupled to gravity we put $n$ scalar fields with 
$n = 0,1,2$ and $3$ and/or Ising spins on each triangle.
\subsection{Resistivity distributions}
We pick-up five vertices randomly as five electrodes to obtain the 
resistivity constant as indicated in Sec. 2, and repeat this procedure 
$50$ times for each configuration, for about $100$ independent configurations 
as ensembles.

\begin{figure}[h]
\vspace{-0.5cm}
\centerline{
\psfig{file=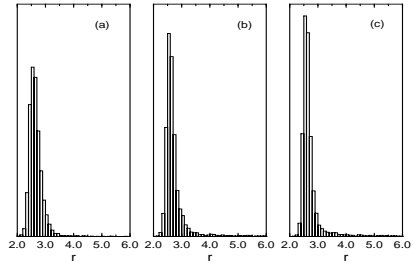,height=3.8cm,width=5.8cm}} 
\vspace{-1.0cm}
\caption
{
Resistivities for the case of pure gravity. 
The size of ($a$), ($b$) and ($c$) are $4000$, $8000$ and $16000$ triangles 
respectively.
}
\label{fig:Pure_Resist}
\vspace{-0.3cm}
\end{figure}

\begin{figure}[h]
\vspace{-1.1cm}
\centerline{
\psfig{file=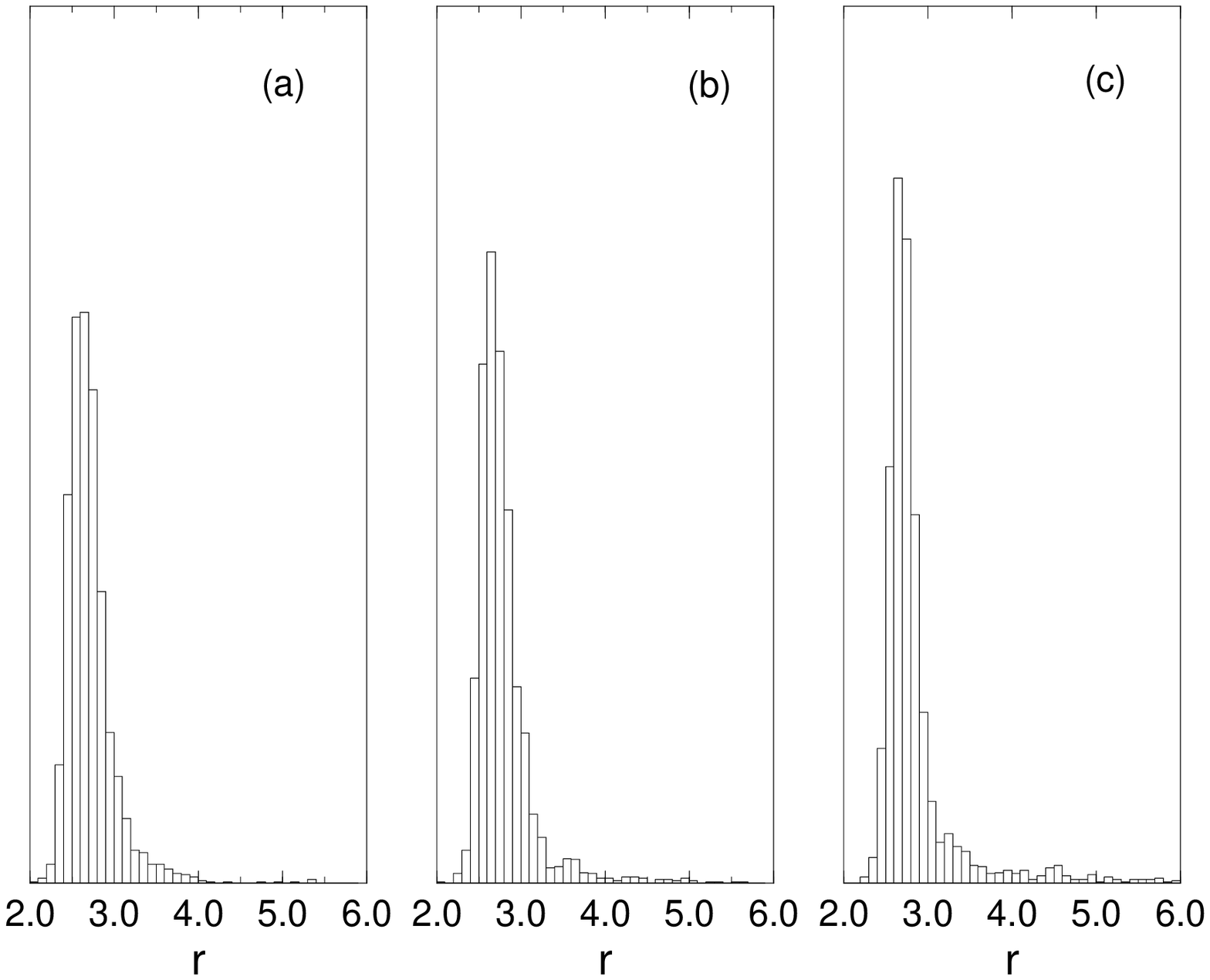,height=3.8cm,width=5.8cm}} 
\vspace{-1.0cm}
\caption
{
Resistivities for the case of 1 scalar.
}
\label{fig:B1_Resist}
\vspace{-0.8cm}
\end{figure}

\begin{figure}[h]
\vspace{0cm}
\centerline{\psfig{file=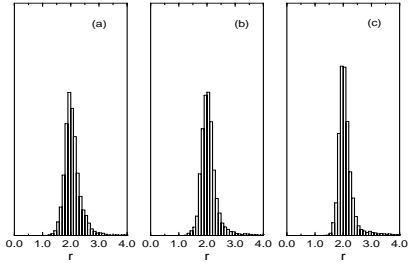,height=3.8cm,width=5.8cm}} 
\vspace{-1.0cm}
\caption
{
Resistivities for the case of pure gravity when each resistance is randomly 
distributed.
}
\label{fig:Random_resist}
\vspace{-0.1cm}
\end{figure}

\begin{figure}[h]
\vspace{-0.1cm}
\centerline{
\psfig{file=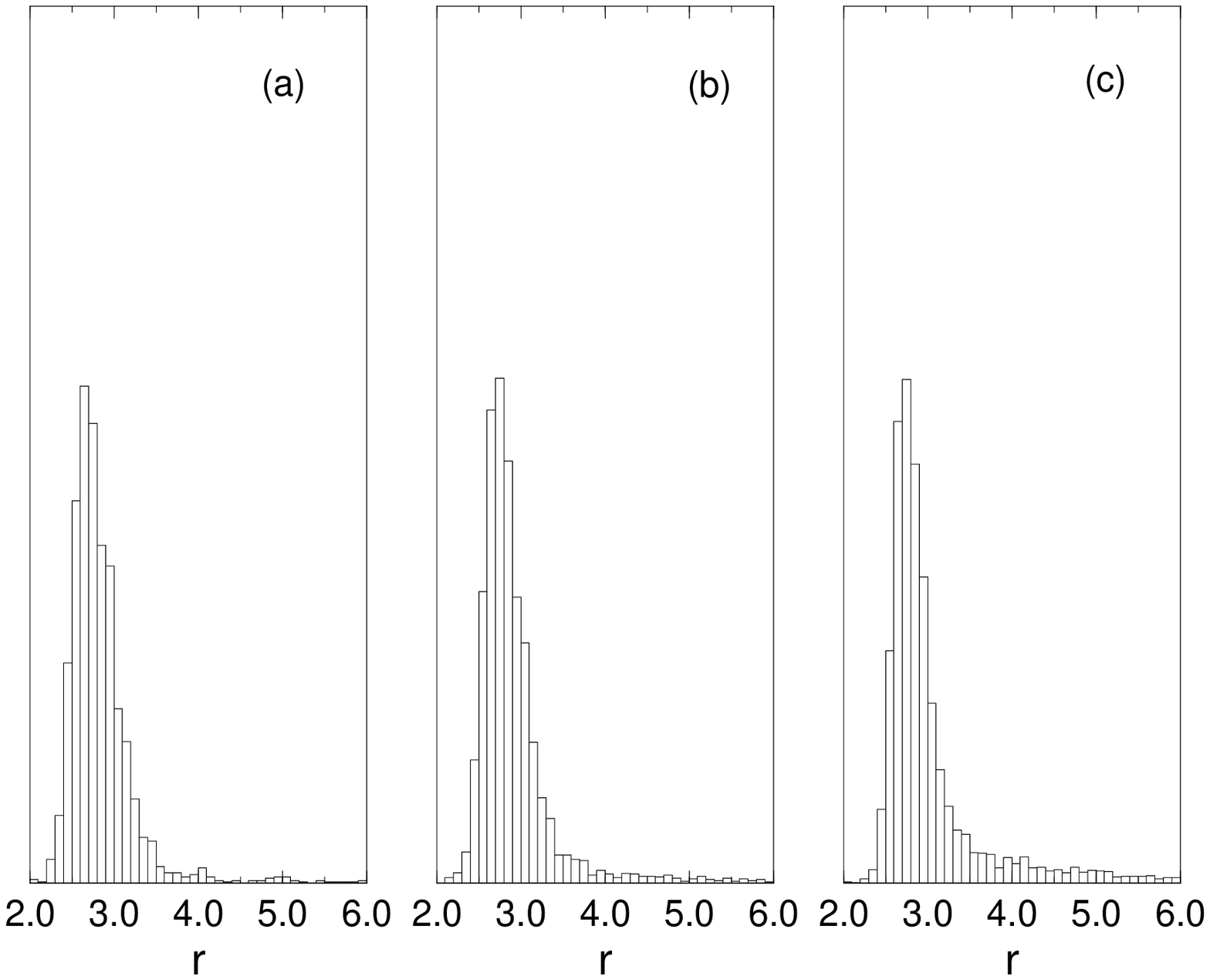,height=3.8cm,width=5.8cm}} 
\vspace{-1.0cm}
\caption
{
Resistivities for the case of 2 scalars.
}
\label{fig:B2_Resist}
\vspace{-0.1cm}
\end{figure}

\begin{figure}[h]
\vspace{-0.1cm}
\centerline{
\psfig{file=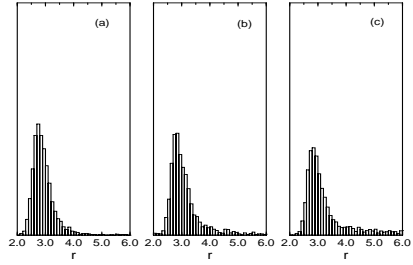,height=3.8cm,width=5.8cm}} 
\vspace{-1.0cm}
\caption
{
Resistivities for the case of 3 scalars.
}
\label{fig:B3_Resist}
\vspace{-0.1cm}
\end{figure}

Let us first take a look at the results for pure gravity
(Fig.\ref{fig:Pure_Resist}).
We find that {\it peaks get narrower as the size grows}.
The value $r \approx 2.6$ should be compared to $\sqrt {3}$ of
the flat network($i.e.$ a honeycomb lattice).
An increase in $r$ is understood to be a reflection of the fractal nature 
of the surface.
As the surfaces get larger, finite size effects due to the discreteness 
diminish and eventually the peak will grow indefinitely. 
This is what we have expected as the continuum limit of a network of resistors.
We find similar tendency in the one scalar case($c=1$)(Fig.\ref{fig:B1_Resist}). 
Also, in Fig.\ref{fig:Random_resist}, we show data to justify the assumptions 
of the analog model for the case of pure gravity.
Each resistance is distributed randomly with weight $e^{-\alpha_{l}}$.
These distributions indicate the same behavior with previous cases $c=0$.
In Fig.\ref{fig:B2_Resist},\ref{fig:B3_Resist} we plot the distributions of 
$r$ for the $c=2,3$ cases.
Here no sharpening of peaks with increasing sizes is seen unlike the previous 
cases, which suggests that the surface does not approach a smooth continuum 
limit. 
This is considered as the reflection of the tachyonic mode instability in the 
continuous theory.
The same tendency is also supported in the case of 3 critical Ising spins.

In conclusion we have observed the so called $c=1$ barrier for two 
dimensional quantum gravity coupled to matter as the transition from a 
well-defined complex structure to an ill-defined one. 


\subsection{Separating the conformal modes}
Next, we discuss how to extract information on the conformal mode.
Once we have found the value of resistivity, we can assign a complex 
coordinate to each vertex of the dual graph from eq.(\ref{eq:potd1}) and 
eq.(\ref{eq:potd2}).
In Fig.6 we plot the distribution of coordinates for all the 
vertices of a surface for the case of pure gravity with $8000$ triangles. 
The point density around a point $z$ should be proportional to 
$\sqrt {g(z)}$, because each vertex is supposed to carry the same 
space-time volume.
In the conformal gauge which we have been employing, we have 
$\sqrt {g(z)} = \exp {\phi(z)}$, where $\phi(z)$ is the conformal mode.
This argument may be over-simplified in the sense that we did not pay 
attention to the quantum fluctuation of $\phi(z)$ and the renormalization 
of the composite operator $\exp {\phi(z)}$.
What we can say safely is that the point density at $z$ after taking an 
ensemble average should be equal to the four-point function 
$$
\lim_{w \to \infty} |w|^{4} < e^{\alpha\phi(0)}e^{\alpha\phi(1)}
e^{\alpha\phi(w)}e^{\alpha\phi(z)} >,
$$ 
in the Liouville theory, where $w$ is a point on the complex plane and 
$\alpha$ is the renormalization factor for the conformal mode.
In a recent paper\cite{Zamol_Zamol}, an estimate for this four-point function 
has been given. This should be compared with 
Fig.6.

\begin{center}
{\Large Acknowledgements}
\end{center}
We are grateful to N.D.Hari Dass and N.Ishibashi for useful discussions 
and comments.
One of the authors(N.T.) is supported by Research Fellowships of the Japan 
Society for the Promotion of Science for Young Scientists.

\end{document}